\documentclass[journal]{IEEEtran}
\usepackage{graphicx}
\ifCLASSINFOpdf

\else

\fi

\begin{document}

\title{Dense-Coding Attack on Three-Party Quantum Key Distribution Protocols}

\author{Fei~Gao, Su-Juan~Qin, Fen-Zhuo Guo, and~Qiao-Yan~Wen
\thanks{All authors are with State Key Laboratory of Networking and Switching
Technology, Beijing University of Posts and Telecommunications,
Beijing 100876, China (e-mail: gaofei\_bupt@hotmail.com;
qsujuan@bupt.edu.cn; gfenzhuo@bupt.edu.cn; wqy@bupt.edu.cn).}}

\markboth{IEEE JOURNAL OF Quantum Electronics,~Vol.~X, No.~X, January~2010}%
{Shell \MakeLowercase{\textit{et al.}}: Bare Demo of IEEEtran.cls
for Journals}

\maketitle

\begin{abstract}
Cryptanalysis is an important branch in the study of cryptography,
including both the classical cryptography and the quantum one. In
this paper we analyze the security of two three-party quantum key
distribution protocols (QKDPs) proposed recently, and point out
that they are susceptible to a simple and effective attack, i.e.
the dense-coding attack. It is shown that the eavesdropper Eve can
totally obtain the session key by sending entangled qubits as the
fake signal to Alice and performing collective measurements after
Alice's encoding. The attack process is just like a dense-coding
communication between Eve and Alice, where a special measurement
basis is employed. Furthermore, this attack does not introduce any
errors to the transmitted information and consequently will not be
discovered by Alice and Bob. The attack strategy is described in
detail and a proof for its correctness is given. At last, the root
of this insecurity and a possible way to improve these protocols
are discussed.
\end{abstract}

\begin{IEEEkeywords}
Quantum cryptography, Quantum key distribution, quantum network
communication, cryptanalysis, dense coding.
\end{IEEEkeywords}

\IEEEpeerreviewmaketitle

\section{Introduction}

\IEEEPARstart{C}{ryptography is}  the approach to protect data
secrecy in public environment. As we know, the security of most
classical cryptosystems is based on the assumption of
computational complexity. But it was shown that this kind of
security might be susceptible to the strong ability of quantum
computation \cite{Shor,Grover}. That is, many existing
cryptosystems will become no longer secure once quantum computer
appears.

Fortunately, this difficulty can be overcome by quantum
cryptography \cite{GRTZ02,TTMTT07}. Different from its classical
counterpart, quantum cryptography is the combination of quantum
mechanics and cryptography, where the security is assured by
physical principles such as Heisenberg uncertainty principle and
quantum no-cloning theorem. Now quantum cryptography has attracted
a great deal of attentions because it can stand against the threat
from an attacker with the ability of quantum computation. Quite a
few branches of quantum cryptography have been studied in recent
years, including quantum key distribution (QKD)
\cite{BB84,I06ie,TMTTT09,Y09ie,MTTTT09,XCWY09}, quantum secret
sharing (QSS) \cite{CGL99,HBB99,KKI99}, quantum secure direct
communication (QSDC) \cite{LL02,BF02,DLL03,CYW10,QWMZ09}, quantum
identity authentication \cite{DHHM99,ZZ2000,Z2009}, and so on.

As the most important application of quantum cryptography, QKD
allows that two users, generally called Alice and Bob, can
privately share a random key by using quantum carriers. The QKD
protocols are designed carefully so that any effective
eavesdropping will result in distortion of the quantum states and
then be discovered by the legal users. The fact that legal users
can discover potential eavesdroppings is the key point of the
security of QKD. If eavesdroppings are detected, the transmitted
key, essentially a sequence of random bits, will be discarded.
Otherwise, a secure key will be shared and it can be employed to
encrypt the secrets communicated between Alice and Bob.

Obviously, quantum devices (QDs) are necessary to realize a QKD
protocol, including the devices to generate qubits, to store
qubits, to measure qubits, to perform unitary operations, and so
on. For example, in the famous BB84 protocol \cite{BB84}, Alice
has to generate qubits in four different quantum states
$|0\rangle$, $|1\rangle$, $|+\rangle$, and $|-\rangle$, while Bob
needs to execute measurements in two different bases
$B_z=\{|0\rangle, |1\rangle\}$ and $B_x=\{|+\rangle, |-\rangle\}$.
Here $|+\rangle=\frac{1}{\sqrt{2}}(|0\rangle+|1\rangle)$,
$|-\rangle=\frac{1}{\sqrt{2}}(|0\rangle-|1\rangle)$. However, QDs
are still expensive because qubits are quite difficult to deal
with. In fact this is also the main reason why quantum
cryptography has not been widely used in our daily life.
Therefore, it is desirable to design protocols where some QDs are
shared by different users. To this aim, a new QKD model, i.e.
three-party QKD \cite{LZWD05,PBTB95,SLH09}, appeared. Till now,
most QKD protocols are two-party ones. That is, only two users,
Alice and Bob, are concerned. In three-party QKD, another
participant, i.e. the center, is introduced to help Alice and Bob
to distribute the key. Furthermore, the center is equipped most of
QDs while the users has less. When this kind of QKD is implemented
in a network, one center can provide service to many users. By
this means expensive QDs are shared and the expense of every user
is reduced. As a result, three-party version is an effective
manner to keep down the cost of QKD.

As we know, design and analysis has always been important branches
of cryptography. Both of them drive the development of this field.
In fact, cryptanalysis is an important and interesting work in
quantum cryptography. As pointed out by Lo and Ko, \emph{breaking
cryptographic systems was as important as building them}
\cite{LK05}. In a QKD protocol, it is generally supposed that the
quantum channel can be attacked with any manner allowed by quantum
mechanics, while the classical one can only be listened but cannot
be modified \cite{GRTZ02,BB84}. In this situation, we say an
attack strategy is successful if the eavesdropper Eve can elicit
all or part of the secret key without being discovered by Alice
and Bob.

Though in quantum cryptography legal users generally have the
ability of discovering potential eavesdroppings, not all proposed
protocols can achieve their expected security. Some protocols were
attacked successfully by subtle strategies which were not
concerned when these protocols were originally designed. Quite a
few effective attack strategies have been proposed, such as
intercept-resend attack \cite{GGW08}, entanglement-swapping attack
\cite{ZLG01}, teleportation attack \cite{GGWcpb08,Goc09,ZZ10cpl},
channel-loss attack \cite{W03,W04}, Denial-of-Service (DoS) attack
\cite{C03,GGWZpra08}, Correlation-Extractability (CE) attack
\cite{GWZpla06,GLWZcpl08,GQWZoc10}, Trojan horse attack
\cite{GFKZ06,DLZZ05}, participant attack \cite{GQW07,QGW06,HLL10},
and so on. Understanding those attacks will be helpful for us to
design new schemes with high security.

Recently Shih, Lee, and Hwang presented two novel three-party QKD
protocols \cite{SLH09}, where one is executed with an honest
center and the other is with an untrusted center. Here ``honest''
means the center always follows the designed procedures to execute
the protocol, and ``untrusted'' implies the center might cheat
Alice and Bob, and try to elicit the key like an attacker. In this
paper, we analyze the security of these two three-party QKD
protocols and show that Eve can obtain the whole key transmitted
between Alice and Bob without being detected by legal users. This
attack is based on the technique of dense coding \cite{BW92},
which was also used in pervious strategies \cite{QGW06,HLL10}.

The rest of this paper is organized as follows. The next section
describes the three-party QKD protocols in Ref.~\cite{SLH09} and
introduces dense coding. In section III the particular attack
strategy is demonstrated, and some useful discussions, including
how to improve the protocols, are given in Section IV. Finally, a
short conclusion is given in Section V.

\section{Previous works}
In this section we will describe the three-party QKD protocols
presented in Ref.~\cite{SLH09} and a special feature of quantum
entanglement, i.e. dense coding, which is useful in our attack
strategy.

\subsection{The three-party QKD protocols}
In Ref.~\cite{SLH09} two three-party QKD protocols were proposed.
One deals with an honest center and the other is with an untrusted
center. In the following description, for the sake of simplicity,
we will call these two protocols QKDP-I and QKDP-II, respectively,
and use the same notations as that in Ref.~\cite{SLH09}.

\begin{figure}
\begin{center}
\includegraphics{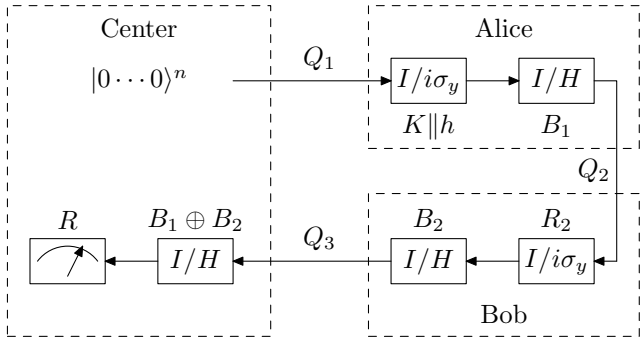}
\caption{\label{fig:one} The process of QKDP-I. For the sake of
simplicity, all classical communications are omitted.}
\end{center}
\end{figure}

Now let us see QKDP-I first, where the technique of ``block
transmission'', proposed in Ref. \cite{LL02}, is utilized. This
protocol is composed with the following steps (see Fig.1).

1. The center generates $n$ qubits $|0\rangle$ and sends this
sequence (denoted as $Q_1$) to Alice.

2. After receiving $Q_1$, Alice selects a $u$-bit random session
key $K$ and computes its $m$-bit hash value $h=H(K)$ as the
checksum, where $u+m=n$. Then Alice performs unitary operation
$U_0=I$ ($U_1=i\sigma_y$) on the $i$-th qubit ($1\leq i\leq n$) in
$Q_1$ if the $i$-th bit in $K\|h$ is 0 (1). Furthermore, Alice
generates an $n$-bit random string $B_1$, and performs unitary
operation $U_0=I$ ($U_2=H$) on the $i$-th qubit in $Q_1$ if the
$i$-th bit in $B_1$ is 0 (1). After these coding operations Alice
sends the new qubit sequence (denoted as $Q_2$) to Bob. Here
\begin{eqnarray}
I=\left[ \begin{array}{c c} 1 & 0 \\
0 & 1
\end{array} \right],
i\sigma_y=\left[ \begin{array}{c c} 0 & 1 \\
-1 & 0
\end{array} \right],
H=\frac{1}{\sqrt{2}}\left[ \begin{array}{c c} 1 & 1 \\
1 & -1
\end{array} \right].
\end{eqnarray}

3. After receiving $Q_2$, Bob selects two $n$-bit random strings
$R_2$ and $B_2$. Then he performs unitary operation $U_0$ or $U_1$
on each qubit in $Q_2$ according to $R_2$, and then operation
$U_0$ or $U_2$ on each qubit according to $B_2$. These coding
operations are similar to Alice's in the previous step. Afterwards
Bob sends the new qubit sequence (denoted as $Q_3$) to the center.

4. The center informs Alice and Bob after the receiving of $Q_3$.

5. Alice and Bob tell the center $B_1$ and $B_2$ respectively.

6. According to $B_1\oplus B_2$, the center recovers the original
polarization bases of qubits by performing $U_0$ or $U_2$ on each
qubit as in steps 2 and 3. Then the center measures all the qubits
in basis $R=\{|0\rangle,|1\rangle\}$, obtaining the measurement
results $C'=R_2\oplus(K\|h)$. At last the center announces $C'$ to
Bob.

7. Bob recovers $K\|h=R_2\oplus C'$ and verifies whether $h=H(K)$.
If it is correct, Bob obtains the session key $K$ and tells Alice
it is successful.

This is the end of QKDP-I. In addition, Alice and Bob would also
take some measures to prevent Trojan horse attack. In this
protocol the operations of qubit generation and measurement are
focused in the center's lab, and Alice and Bob only need to
perform unitary operations on the qubits. As analyzed in
Ref.~\cite{SLH09}, this protocol has high efficiency.

It is easy to see that the center can obtain the session key in
QKDP-I if he/she is not honest \cite{SLH09}. QKDP-II can resolve
this problem, which is suitable for the situation where the center
is untrusted. Now we introduce QKDP-II in brief (see Fig.2), which
is useful when we discuss how to improve the protocols in section
IV. The first two steps are the same as that in QKDP-I. After Bob
received $Q_2$, Alice tells Bob the value of $B_1$. According to
$B_1$, Bob performs $I$ or $H$ to recover the original
polarization bases of the qubits. Then Bob shuffles the sequence
of qubits and sends it (denoted as $Q_3$) to the center. Here the
shuffle operation is actually the technique of ``order
rearrangement'' proposed in Ref. \cite{DL2003}. When the center
receives $Q_3$, he/she measures all qubits in basis $R$, and
announces the results $C'=shuffled\_(K\|h)$ to Bob. At last Bob
rearranges the string $C'$ to obtain $K\|h$, and checks whether
$h=H(K)$. If it is correct, Bob tells Alice it is successful.

\begin{figure}
\begin{center}
\includegraphics{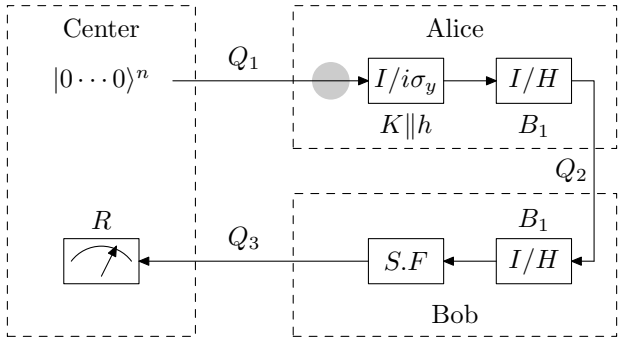}
\caption{\label{fig:two} The process of QKDP-II. The classical
communications are omitted. $S.F$ denotes the shuffle operation,
and the gray circle represents the position where another shuffle
operation should be added to improve the protocol.}
\end{center}
\end{figure}

\subsection{Dense coding}

In 1992 C. H. Bennett and S. J. Wiesner presented a special
feature of Einstein-Podolsky-Rosen (EPR) states, i.e. dense coding
\cite{BW92}. It was shown that two bits of classical information
can be encoded into an EPR state by one-particle unitary
operations. Specifically, if Alice and Bob hold one particle from
an EPR state respectively, Alice can send two bits to Bob by
performing one of four unitary operations on her particle and
transmitting it to Bob. One particle carries two bits of
information, which is the reason why it is called dense coding.
Now we describe how it happens in brief.

Four EPR states are
\begin{eqnarray}
|\Phi^\pm\rangle_{12}=\frac{1}{\sqrt{2}}(|00\rangle\pm|11\rangle)_{12}, \nonumber \\
|\Psi^\pm\rangle_{12}=\frac{1}{\sqrt{2}}(|01\rangle\pm|10\rangle)_{12},
\end{eqnarray}
where the subscripts 1 and 2 denote different particles. These
states are orthogonal with each other and compose a complete
basis, i.e Bell basis $B_{Bell}$. There are also four one-particle
unitary operations $I$, $\sigma_x$, $i\sigma_y$, and $\sigma_z$,
where
\begin{eqnarray}
\sigma_x=\left[ \begin{array}{c c} 0 & 1 \\
1 & 0
\end{array} \right], \hspace {4mm}
\sigma_z=\left[ \begin{array}{c c} 1 & 0 \\0 & -1
\end{array} \right].
\end{eqnarray}

Without loss of generality, suppose Alice and Bob share an EPR
state $|\Phi^+\rangle_{12}$, that is, Alice has particle 1 and Bob
holds 2. Alice can encode two bits of information into the state
by performing one of the above four operations on particle 1,
under which this state changes as
\begin{eqnarray}
I^1|\Phi^+\rangle_{12}=|\Phi^+\rangle_{12}, \hspace{4mm}
\sigma_x^1|\Phi^+\rangle_{12}=|\Psi^+\rangle_{12}, \nonumber \\
(i\sigma_y)^1|\Phi^+\rangle_{12}=|\Psi^-\rangle_{12}, \hspace{4mm}
\sigma_z^1|\Phi^+\rangle_{12}=|\Phi^-\rangle_{12},
\end{eqnarray}
where the superscript represents the qubit on which the operations
are performed. Afterwards Alice sends particle 1 to Bob. Bob can
distinguish which operation is chosen by Alice via a Bell
measurement on particles 1 and 2. If $I$, $\sigma_x$, $i\sigma_y$,
and $\sigma_z$ represent 00, 01, 10, and 11 respectively, Bob can
obtain two bits from Alice. For example, Bob knows Alice's message
is 10 if his measurement result is $|\Psi^-\rangle_{12}$.
Similarly, any one of the four EPR states can be used as the
original state in this communication.

In fact, the above dense coding can be generalized to that using
other entangled states and operations. In the following we will
utilize this idea to design an effective attack on QKDP-I and
QKDP-II, where Eve can totally obtain the transmitted key without
being discovered.

\section{Dense-coding attack}

We take QKDP-I as our example to analyze its security. In this
protocol, Alice encodes the session key $K$ into the qubits in
$Q_1$ by unitary operations $I$ and $i\sigma_y$. To prevent Eve
from obtaining $K$ from these qubits, Alice randomly changes the
basis of each qubit by $I$ and $H$. After the above two operations
every qubit is randomly in one of four nonorthogonal states
$\{|0\rangle,|1\rangle,|+\rangle,|-\rangle\}$. Then Alice sends
the new sequence $Q_2$ to Bob. If Eve intercepts $Q_2$ and wants
to obtain $K$ by measurements, she cannot distinguish the above
four states with certainty and disturbance will be inevitably
introduced to the quantum states. This is very similar with that
in BB84 QKD protocol \cite{BB84}.

However, as we know, fake-signal attack \cite{QGW06} is very
common in the analysis of quantum cryptography. If Eve replaces
the qubits in $Q_1$ by the ones from some entangled states, can
she obtain $K$ by collective measurements after Alice encoded and
sent them out? This question is very interesting and difficult to
give an answer. In fact, the answer is yes. This attack just like
that Alice sends the secret key to Eve by an ``unnoticed dense
coding''. In the following we will depict this attack first and
then prove its correctness.
\begin{figure}
\begin{center}
\includegraphics{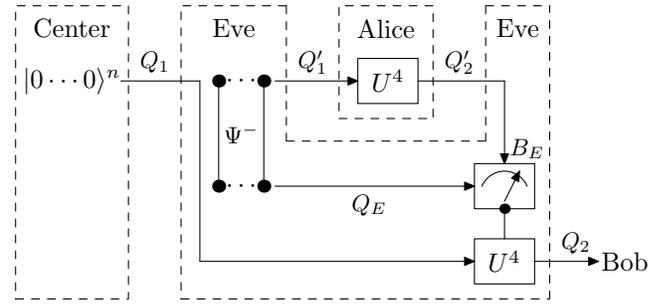}
\caption{\label{fig:three} Dense-coding attack on QKDP-I. The
classical communications are omitted. $U^4$ represents one of the
four operations $\{I,i\sigma_y,H,Hi\sigma_y\}$.}
\end{center}
\end{figure}

Eve's dense-coding attack is as follows (see Fig.3).

E1. Eve generates $n$ ordered EPR pairs in the state
$|\Psi^-\rangle_{12}$. All the qubits with subscript 1 (2) compose
a qubit sequence $Q'_1$ ($Q_E$).

E2. When the center sends the sequence $Q_1$ to Alice in Step 1,
Eve intercepts all the qubits and replaces them by the sequence
$Q'_1$.

E3. When Alice sends the sequence $Q'_2$, i.e. the qubits after
Alice's encoding operations, to Bob in Step 2, Eve intercepts it
and performs collective measurements on every pair of the
corresponding qubits in $Q'_2$ and $Q_E$ in the basis
$B_E=\{|\Psi^-\rangle,|\Phi^+\rangle,|\Omega\rangle,|\Gamma\rangle\}$.
Here
\begin{eqnarray}
|\Omega\rangle_{12}=\frac{1}{2}(|00\rangle-|01\rangle-|10\rangle-|11\rangle)_{12}, \nonumber \\
|\Gamma\rangle_{12}=\frac{1}{2}(|00\rangle+|01\rangle+|10\rangle-|11\rangle)_{12},
\end{eqnarray}
and it is easy to verify that $|\Psi^-\rangle$, $|\Phi^+\rangle$,
$|\Omega\rangle$, and $|\Gamma\rangle$ are orthogonal with each
other.

E4. Eve performs one of the four operations
$\{I,i\sigma_y,H,Hi\sigma_y\}$ on every legal qubits in the
sequence $Q_1$ according to her measurement result. In particular,
Eve performs $I,i\sigma_y,H,Hi\sigma_y$ on the $i$-th qubit in
$Q_1$ when her measurement result on the $i$-th pair is
$|\Psi^-\rangle,|\Phi^+\rangle,|\Omega\rangle,|\Gamma\rangle$,
respectively. Then Eve sends the new sequence $Q_2$, i.e. the
legal qubits after Eve's operations, to Bob.

E5. Eve obtains $K\|h$ from her measurement results in Step E3.
That is, she knows the $i$-th bit of $K\|h$ is 0 (1) if her
measurement result on the $i$-th pair is $|\Psi^-\rangle$ or
$|\Omega\rangle$ ($|\Phi^+\rangle$ or $|\Gamma\rangle$).

By this strategy, Eve will obtain the key $K$ correctly if we do
not consider the errors brought by channel noise or other
eavesdroppings. Furthermore, as we will show, this attack will not
be discovered by legal users. Therefore, it is very effective
though it looks quite simple.

Now we prove the correctness of our attack on QKDP-I. In fact,
Alice's two encoding operations $I/i\sigma_y$ and $I/H$ can be
treated as one operation, i.e. one of
$\{I,i\sigma_y,H,Hi\sigma_y\}$. If Eve can distinguish which one
of the four operations is used by Alice in her encoding, Eve will
obtain not only $K\|h$ but also $B_1$. This situation is quite
similar to that of dense coding. It is not difficult to verify
that the four states in Eve's measurement basis $B_E$ satisfies
\begin{eqnarray}
|\Psi^-\rangle_{12}=I^1|\Psi^-\rangle_{12},~|\Phi^+\rangle_{12}=(i\sigma_y)^1|\Psi^-\rangle_{12},\nonumber \\
|\Omega\rangle_{12}=H^1|\Psi^-\rangle_{12},~|\Gamma\rangle_{12}=(Hi\sigma_y)^1|\Psi^-\rangle_{12}.
\end{eqnarray}
Therefore, when Alice performs her encoding operations on the fake
qubits in $Q'_1$, the quantum state changes as described in Tab.1.

\vspace{10pt} {\small Tab.1. State changes after Alice's encoding
on the fake qubits. The first and the last columns are the
original and the final states of Eve's entangled pairs,
respectively. The second and third columns are bit values of
$K\|h$ and $B_1$, respectively. The fourth column is the combined
operation for Alice's encoding. \vspace{3pt}
\begin{center}
\renewcommand\arraystretch{2.0}
\begin{tabular}{ccccc} \hline
Orig. Stat. & $k\|h$ & $B_1$ & Comb. Oper. & Fina. Stat.\\ \hline
$|\Psi^-\rangle$ & 0 & 0 & $I$ & $|\Psi^-\rangle$ \\
$|\Psi^-\rangle$ & 0 & 1 & $H$ & $|\Omega\rangle$ \\
$|\Psi^-\rangle$ & 1 & 0 & $i\sigma_y$ & $|\Phi^+\rangle$ \\
$|\Psi^-\rangle$ & 1 & 1 & $Hi\sigma_y$ & $|\Gamma\rangle$\\\hline
\end{tabular}
\end{center}}
\vspace{10pt}

Consider the $i$-th pair of qubits as our example, which is
originally generated by Eve in the state $|\Psi^-\rangle_{12}$.
When Eve sends the first qubit to Alice in Step E2, Alice will
encode the $i$-th bits of $K\|h$ and $B_1$ on it. Without loss of
generality, if the $i$-th bits of $K\|h$ and $B_1$ are 1 and 0
respectively, as shown in the fourth row in Tab.1, Alice's
combined operation will be $i\sigma_y$, and then the state will be
changed into $|\Phi^+\rangle$ after the encoding.

It can be seen from Tab.1 that four possible final states include
$|\Psi^-\rangle$, $|\Phi^+\rangle$, $|\Omega\rangle$, and
$|\Gamma\rangle$, which are orthogonal with each other and can be
distinguished with certainty by measurements in basis $B_z$.
Therefore, Eve's measurement results $|\Psi^-\rangle$,
$|\Phi^+\rangle$, $|\Omega\rangle$, and $|\Gamma\rangle$ on the
$i$-th pair imply that the $i$-th bits of $K\|h$ and $B_1$ are 00,
10, 01, and 11, respectively. As a result, Eve can get the correct
session key in Step E5.

Obviously, with the knowledge that which operations has been
chosen by Alice, Eve just performs the same operations on the
legal qubits in $Q_1$ in step E4. Therefore, the states of the new
sequence $Q_2$ is the same as the situation where no eavesdropping
happens. Consequently, no errors will be introduced by this attack
and Eve will never be discovered. Note that every fake signal Eve
sends to Alice only contains one ordinary qubit, which is
different from a spy photon or invisible photon, and hence it
would be unnoticed by Alice's apparatus, including that to prevent
Trojan horse attack.

In a word, the dense-coding attack is correct and very effective
for QKDP-I. Additionally, this attack is also suitable for QKDP-II
because there is no difference between both protocols when Alice's
encoding operations are considered (see Alice's areas in Fig.1 and
Fig.2).

\section{Discussions}
Now we give some discussions about the security of two three-party
QKDPs and our attack.

As we all know, fake-signal attack is very common in quantum
cryptography and some effective manners to prevent it have been
found. Then a question arises, i.e., why the three-party QKDPs are
susceptible to such a familiar attack? In fact quite a few
protocols \cite{DL04pra,LM05prl,DZL05pla} use similar properties
of single photons, including the carrier states and the encoding
operations, but they are all secure against this kind of attack.
By careful comparison between the three-party QKDPs and these
secure ones we can find the answer of the above question. That is,
in the secure protocols the users will detect eavesdropping by
some manners such as conjugate-basis measurements after he/she
received the qubits, while it does not happen in the three-party
QKDPs. Obviously, without any detections, Alice can never discover
that the qubits were replaced by Eve when they were transmitted in
public channel. In the three-party QKDPs, to reduce the cost of
users, Alice and Bob have no measurement apparatus, and
consequently cannot take general measures to detect eavesdropping.
Though Bob checks whether $h=H(K)$ at last, which is the only
detection in the three-party QKDPs, it is not strong enough.
Therefore, it should be emphasized that more attention should be
paid to the protocol's security when we pursues low expenses for
the users because low expenses generally implies low capability to
detect eavesdropping.

Now we discuss how to improve the three-party QKDPs to stand
against the dense-coding attack. Note that in QKDP-II the
operation of shuffle is utilized to prevent a dishonest center
from obtaining the key. Actually this technique can also be used
to protect the protocol against Eve's dense-coding attack. If
Alice adds a shuffle operation before her encodings in QKDP-II
(the position is denoted by the gray circle in Fig.2), Eve's
attack will be of no effect. On the one hand, after Alice's
encoding every qubit in $Q'_2$ is in the same state, i.e. the
maximally mixed state $\rho=I$, and then Eve cannot distinguish
them and know which two qubits are originally in an entangled
pair. Therefore, the measurement result will be random if Eve
still measures every two qubits in the same position in $Q'_2$ and
$Q_E$. Thus, Eve cannot distill the key from her measurement
results and the eavesdropping will be discovered when Bob checks
whether $h=H(K)$. On the other hand, because all qubits in $Q_1$
are in the same state $|0\rangle$, the additional shuffle brings
no changes for legal qubits. As a result, this simple modification
is very effective and interesting in the sense that it can prevent
the dense-coding attack but has no effects to the protocol when no
eavesdropping happens.

As shown above, the three-party QKDPs are insecure under the
dense-coding attack. One may want to know what is wrong with the
security proof in Ref.~\cite{SLH09}. In fact the authors of
Ref.~\cite{SLH09} presented a formal proof with the
sequence-of-games approach, which is often used in classical
cryptography \cite{S06ol,N07ics,BSZ07}. However, the attack
strategies considered in this proof are not complete and the
dense-coding attack is overlooked. As we know, quantum mechanics
have many interesting or even counterfactual features. They not
only give convenience for the users to distribute a secret key but
also bring different kinds of new attack strategies for the
eavesdropper. Therefore, more attention should be paid to all
kinds of possible attacks in analyzing the security of a quantum
cryptographic protocols.

\section{Conclusion}
Recently, two novel three-party QKDPs were proposed, in which, to
save the expense, qubit generation and measurement are not needed
for Alice and Bob \cite{SLH09}. In this paper we analyze the
security of these protocols and find that they are susceptible to
a special attack, i.e. the dense-coding attack. In this attack the
eavesdropper Eve can obtain all the session key by sending
entangled qubits to Alice and performing collective measurements
after Alice's encoding, which is just like the process of dense
coding between Eve and Alice. Furthermore, this attack does not
introduce any errors to the transmitted information and
consequently will not be discovered by Alice and Bob. The attack
strategy is described in detail and a proof for its correctness is
given. At last, a possible way to improve these protocols is
discussed.

\section*{Acknowledgment}

This work is supported by NSFC (Grant Nos. 60873191, 60903152,
61003286, 60821001), SRFDP (Grant Nos. 200800131016,
20090005110010), Beijing Nova Program (Grant No. 2008B51), Key
Project of Chinese Ministry of Education (Grant No. 109014).

\ifCLASSOPTIONcaptionsoff
  \newpage
\fi



%

\end{document}